\documentclass[a4paper,12pt,nohyper]{JHEP3}

\newcommand{\bea}{\begin{eqnarray}}
\newcommand{\eea}{\end{eqnarray}} 
\newcommand{\be}{\begin{equation}} 
\newcommand{\bel}[1]{\begin{equation}\label{#1}}
\newcommand{\ee}{\end{equation}} 
\newcommand{\rf}[1]{(\ref{#1})}

\newcommand{\bit}{\begin{itemize}}
\newcommand{\eit}{\end{itemize}}
\newcommand{\ben}{\begin{enumerate}}
\newcommand{\een}{\end{enumerate}}

\def\del{\partial}

\def\alp{\leavevmode\ifmmode {\alpha^\prime} \else ${\alpha^\prime}$ \fi}

\title{Winding Strings in Singular Spacetimes}

\preprint{}

\author{Marek Paw{\l}owski, W{\l}odzimierz Piechocki and Micha{\l} Spali\'nski  
\\ \\
So\l tan Institute for Nuclear Studies\\ 
ul. Ho\.za 69, 
00-681 Warszawa, Poland.
}

\abstract{

Evolution of winding strings in spacetimes with cycles whose
proper lengths depend on time is examined. It was established earlier that
extended objects wrapping the shrinking dimension in compactified Milne
spacetime enjoy classically nonsingular evolution.  Extensions of this
observation to other spacetimes are discussed.
}


\begin{document}

\setcounter{tocdepth}{2}

\section{Introduction}

One of the most fascinating questions in cosmology concerns the nature of
the apparent singularity assumed to lie at the origin of the Big Bang. 
For many reasons this question cannot be resolved within classical general
relativity, and it is natural to seek insights based on string theory,
which has provided numerous examples of consistent physics even in
background spacetimes which are classically singular. Perturbative examples
of this include, for example, Euclidean orbifold
compactifications \cite{Dixon:1985jw}.  
This has fueled hopes that the initial cosmological singularity
would also find a 
consistent description, perhaps involving a smooth transition from what has
been termed a ``pre-Big-Bang'' era \cite{Gasperini:2002bn}. 
In recent years a lot of attention was devoted to studies of string theory 
spacetimes with
cosmologically relevant features such as time-dependence and occurrence of
space-like singularities. 
These studies focused mostly on 
various 
time-dependent orbifolds 
\cite{Horowitz:ap}-\cite{Berkooz:2005ym} 
(the subject is reviewed in references \cite{Cornalba:2003kd} and
 \cite{Durin:2005ix}).  
This activity was hampered by 
the lack of a proper second-quantized description of string theory, but 
some semi-heuristic approaches to this have been
formulated \cite{Gubser:2003vk}. 

One of the simplest examples of a spacetime background with a space-like
singularity is compactified Milne space \cite{Khoury:2001bz}. Interest in
this particular case 
is motivated by its apparent simplicity, as well as its possible
significance for the cyclic Universe
scenario \cite{Turok:2004yx,Steinhardt:2004gk} (as realized in the 
heterotic M-theory framework \cite{Horava:1996ma,Horava:1995qa}). This 
spacetime involves a compact direction 
whose radius decreases linearly with time $t$ until it reaches zero, and
then 
expands linearly at the same rate.  Point particles, as well as generic
configurations of extended
objects, experience blue-shifting as the radius shrinks. This effect
disappears if the momentum in the shrinking direction vanishes, but even
for such special initial conditions the problem cannot be dismissed, since
any infinitesimal momentum in the shrinking direction will lead to
nonanalytic behavior of generic metric perturbations \cite{Turok:2004gb}
as $t$ goes to zero. Turok et al. \cite{Turok:2004gb} argue that for those
states one has to 
seek a description based on an expansion in inverse powers of \alp,
i.e. opposite to the expansion which leads to Einstein's gravity at large
distances. One may adopt also a somewhat simple-minded view according to
which the blue-shifted objects decouple leaving an effective theory valid
close to the singularity which now involves only states corresponding to
objects whose mass remains finite in that region.
The authors of \cite{Turok:2004gb} 
point out however 
that in contrast to point particles, states of
extended objects winding uniformly around the shrinking direction evolve
classically in a  
smooth and unambiguous way through the singularity. This is due to the fact
that for a wrapped object the dimension which is wrapped ceases to be a
direction of possible physical motion.
The main focus of
 \cite{Turok:2004gb} is actually on winding states of M2-branes in M-theory,
but the 
case of winding strings is interesting in its own right. In the M-theory
context such states arise as M2-branes wrapping tori, whose moduli are
time-dependent. 
Zero-modes of strings wrapped on the Milne circle can be viewed either 
as particles with time-dependent mass in Minkowski space, or as 
particles of fixed mass in a spacetime geometry which is singular due to a
conformal factor which vanishes at $t=0$.  
Unambiguous and smooth evolution of
particles in such a situation is somewhat surprising, and one would 
like to understand better how this comes about. A part of section
\ref{sect:milne} below 
is devoted to this issue.

This note extends the considerations of winding strings and membranes
appearing in \cite{Turok:2004gb} to somewhat more general background
spacetimes involving shrinking cycles. Such backgrounds are of interest to
cosmology, since there is no 
reason to assume that only one compact dimension undergoes periodic
expansion and contraction, as in the simplest incarnations of the cyclic
universe scenario. Indeed, such spacetimes have recently been the focus of
much activity aimed at understanding the current phase of accelerated
expansion of the
universe \cite{Townsend:2003fx}-\cite{Ohta:2004wk}.  
It appears however that the observation of Turok et al. does not directly
generalize to cases when more than one cycle is shrinking. The reasons for
this are discussed in section \ref{sect:multiple}. 

The considerations reported below are presented in terms of bosonic string
theory. More realistically one would need to embed this in a superstring
setting, which involves additional degrees of freedom on the worldsheet. In
particular, choices need to be made as to the boundary conditions for
spacetime fermions when they are continued around any cycles. The mass and
interpretation of quantum string states corresponding to 
the classical winding modes discussed here will depend 
on how this is done. 
This note does not address this issue; the considerations presented here
are classical, in the spirit of \cite{Turok:2004gb}. 
It should also be kept in mind that considerations of time-dependent
backgrounds of  
this type require verifying 
consistency at the quantum level (which for strings implies checking
beta-functions for the 
worldsheet sigma model). Examples are known where this can be done, but
this aspect of the problem is left for future work. 

The focus on winding modes in a cosmological setting has been a leitmotif
in string cosmology since the early papers by Brandenberger and
Vafa \cite{Brandenberger:1988aj}. Quite 
recently the condensation of light winding states in a context very close
to the one discussed here appeared in considerations of chronology
protection \cite{Costa:2005ej}. 

While this note was being written a couple of papers
 \cite{Tolley:2005us,McGreevy:2005ci} appeared, 
where 
similar issues are addressed.

\section{Strings in Milne Spacetime}
\label{sect:milne}

This section is devoted to the example of compactified Milne space
discussed in 
 \cite{Turok:2004gb}. Compactified Milne spacetime has the metric
\be\label{milmetr}
ds^2 = -dt^2 + \beta^2 t^2 d\theta^2 + dx^k dx^k, \qquad k=1\dots d-1 \ ,
\ee
where the coordinate $\theta$ describes an $S^1$, so that 
\be
\theta\approx\theta+2\pi\ ,
\ee
and $\beta t$ is the (instantaneous) radius\footnote{In fact, as
discussed in \cite{Turok:2004gb}, one could just as well have a segment
instead of 
$S^1$ here. The only important point is that this dimension is shrinking as
the 
singularity is approached.} of the $S^1$. The coordinate $t$ is the time,
assumed to run from $-\infty$ to $\infty$. This metric is locally flat, but
it has a curvature
singularity at $t=0$ unless $\beta = 1$. In the latter case the metric
\rf{milmetr} can be brought to flat Minkowski form (globally) by a change
of coordinates. 

The classical dynamics of a relativistic string in a background metric $G$
can be described by means of the Nambu-Goto action \cite{luth,jpol}
\bel{nambugoto}
S = -\frac{1}{4\pi\alp} \int d\tau d\sigma 
\sqrt{(\del_\tau X \cdot \del_\sigma X)^2 -
  (\del_\tau X\cdot\del_\tau X)   (\del_\sigma X\cdot\del_\sigma X) }\ ,
\ee
where $X^\mu(\tau, \sigma)$ denote the string embedding and the ``dot''
denotes the scalar product defined by $G$. Using the notation $(X^0, \dots
, X^d) \equiv (T,X^1,\dots ,X^{d-1},\Theta)$ 
the string zero-modes in the winding sector labeled by $w$ are of the form
\be\label{milnezm}
T = t(\tau),\qquad X^k = x^k(\tau),\qquad \Theta= w\sigma\ .
\ee
The reason why these modes are of special interest was forcefully put forth 
in \cite{Turok:2004gb}. Objects moving in the spacetime \rf{milmetr} will
undergo  
blue-shifting as soon as they have any momentum along the shrinking
direction. Their description becomes invalid as the singularity is
approached. 
The exception are extended objects wrapped around the shrinking
dimension. For such objects motion in this direction is a gauge degree of
freedom, so these objects 
are naturally immune to blue-shifting. Indeed, their mass is to first 
approximation given by their tension times their length, and since their
length tends to zero as the classical singularity is approached, they become
light (and instantaneously massless) as it is crossed. 

The action in the zero-mode sector follows from the Nambu-Goto action 
upon substituting the metric \rf{milmetr} and the ansatz \rf{milnezm}. It
is given by
\bel{milnepartact}
S_0 = -\int d\tau\ m(t) \sqrt{\dot{t}^2-\dot{\vec{x}}^2} \ ,
\ee
where $\vec{x}$ denotes the transverse coordinates $x^1\dots x^{d-1}$,
the dot denotes a derivative with respect to $\tau$, and   
\bel{milmass}
m(t) = \frac{|w t|\beta}{2\alp} \equiv m_0 |t| \ .
\ee
This can be interpreted as the action for a particle in Minkowski space,
with a mass depending on time according to \rf{milmass}. 

It is convenient to fix the gauge $T=\tau$ (synchronous gauge), which
leaves 
the transverse coordinates $x^k$ as the independent degrees of
freedom. The action becomes
\bel{milneact}
S_0 = -\int dt\ m(t) \sqrt{1-\dot{\vec{x}}^2} \ .
\ee
The momentum conjugate to $\vec{x}$ 
\be
\vec{p} = \frac{m(t) \dot{\vec{x}}}{\sqrt{1-\dot{\vec{x}}^2}}
\ee
is conserved. The Hamiltonian is simply 
\bel{milham}
H = \sqrt{\vec{p}^2 + m^2(t)} \ .
\ee
As stressed by Turok et al. \cite{Turok:2004gb} this is not singular at
$t=0$.  
The simple-minded argument given above has also been carried through 
in a more formal manner (in the Hamiltonian formalism) leading to the same
conclusions \cite{Turok:2004gb}.  
The equations of motion 
\bel{eom}
\frac{d\vec{x}}{dt} = \frac{\vec{p}}{\sqrt{\vec{p}^2 + m(t)^2}}
\ee
are solved uniquely by
\bel{asinh}
\vec{x} = \vec{x_0} + \frac{\vec{p}}{m_0}
\ \mathrm{asinh}(\frac{m_0}{|\vec{p}|}t)  \ ,
\ee 
where $\vec{x}_0$ is an integration constant \cite{Turok:2004gb}. 

One can also find this solution starting with the Polyakov form of the 
action in conformal gauge \cite{luth,jpol}:
\bel{polyakov}
S = -\frac{1}{4\pi\alp} \int d\sigma d\tau 
( - \del_+ T \del_- T + T^2 (\del_+\Theta \del_-\Theta) +
\del_+\vec{X}\del_-\vec{X})
\ee
and imposing Virasoro constraints $T_{++} = T_{--} = 0$, where
\bel{stressten}
T_{\pm\pm} =  - \del_\pm T \del_\pm T + T^2 (\del_\pm\Theta \del_\pm\Theta)
+ \del_\pm\vec{X}\del_\pm\vec{X}
\ee
are the nontrivial components of the worldsheet energy-momentum tensor in
lightfront coordinates
$\sigma^\pm\equiv\tau\pm\sigma$. 

There is also an alternative way of interpreting the action
\rf{milnepartact}: 
one can regard it as the action for a particle with fixed
mass in a spacetime with metric
\be\label{redmilne}
ds^2 = t^2 (-dt^2 + dx^k dx^k) \ .
\ee
This metric is singular: its determinant vanishes at $t=0$. Invariants 
built from the Riemann tensor also diverge there, so from the point of view
of differential geometry this is a real singularity, rather than a
coordinate one. 

It is illuminating to write the metric \rf{redmilne} in terms of a
different time coordinate, namely $|\xi|=t^2/2$: 
\be\label{radia}
 ds^2 = -d\xi^2 + 2 |\xi| dx^k dx^k \ .
\ee
If $\xi$ is positive this is the Friedman-Robertson-Walker 
metric for a universe filled with
radiation and $\xi$ would be identified with the cosmic time ($t$ being the 
conformal time). Allowing negative cosmic times in \rf{radia} is an
extension of the standard Friedman-Robertson-Walker  spacetime, but in
contrast to the usual  
sense of extending the domain of validity of the coordinate system, it 
does not remove singularity at $\xi=0$ ($t=0$) which is a real singularity 
of the curvature. In spite of this fact, one is allowed to consider time
like geodesics describing smooth evolution of massive particles through the
singular locus in the extended Friedman-Robertson-Walker spacetime. To
understand this point  
one has to consider the interpretation of the $\xi=0$ locus more
carefully. The metric becomes singular as $\xi$ goes to zero. In the
standard  
Friedman-Robertson-Walker models the singularity in the past is generally
interpreted as a 
signal that space-time has originated from a 
point\footnote{The locus $\xi = 0$ is formally spanned by a three-dimensional 
hyperspace in the space of coordinates, but the singularity of the metric
at 
$\xi=0$ implies that the lengths of all formal curves on this hyperspace
are equal to zero. It could then be argued that this null hyperspace 
physically is just a point.}. In fact, as long as one does not need to 
assume that the set $\xi=0$ has some intrinsic structure (even only a 
dimension in a topological sense) any assumption about such an intrinsic  
structure could be rejected. But the apparent lack of need is not a
sufficient argument to definitely exclude such possibility. The
alternatives could 
be studied as well as their possible physical consequences. One may argue
that while the metric has a real singularity, the problem of finding a  
time-like curve of minimum metric length is consistent in spite of this fact
because the length of any time-like curve is perfectly well defined even if
it happens to cross the singular locus. Therefore, assuming that at  
$\xi=0$ there is a null three-dimensional surface rather than a point, the
variational problem is well posed and as has a unique solution \rf{asinh}.   
Within the framework of general relativity one cannot determine how the
singular locus is to be interpreted. Only physical consequences -- if found 
and confirmed -- could justify {\em a posteriori} or reject the above
assumption. But from the string perspective the situation is clear: 
the interpretation of 
the coordinates $x$ is given -- they are global Minkowski coordinates and
the issue of degeneration does not arise. 
At $t = 0$ only the compact dimension shrinks to a point. There is no
singularity of the $d$-dimensional metric\footnote{The original Milne
metric \rf{milmetr} is $d+1$-dimensional.}. At $t = 0$  
there is a $d-1$-dimensional plane with Minkowski metric (locally on the 
plane and its neighborhood). 
The particle states identified with fundamental modes of
winding strings become massless as $t$ approaches zero. Particle masses vary
according to the time-dependent conformal factor. 
Particle dynamics is
formally equivalent to dynamics of constant mass particles following
geodesics in the conformally flat space-time.

Is this a unique, somewhat pathological example? Its stringy origin
clearly indicates that it is not. In the following section 
this question is taken up in a class of spacetimes generalizing
\rf{milmetr}.

\section{Winding states in general cases with a shrinking cycle}
\label{sect:genwind}

This section is devoted to moving beyond the case of compactified Milne
spacetime. 
It is rather clear that the example
reviewed in the previous section is easy to generalize. A similar
picture should be valid whenever there is a cycle which degenerates with
time. Strings wrapped on it will have zero-modes which to an observer
probing distances much larger than the radius of the cycle will appear as
particles with a time-dependent mass. This argument should be valid at
least in an appropriate adiabatic limit.

To be more explicit, let us focus on metrics which admit an isometry. 
By a
choice of coordinates one can always express such a metric as 
\be\label{kkansatz}
ds^2 = e^{2\phi} (du + A)^2 + ds_\perp^2 \ ,
\ee
where 
\be
ds_\perp^2 = h_{\mu\nu} dx^\mu dx^\nu, \qquad A = A_\mu dx^\mu
\ee
and ${x^\mu, u}$ are the spacetime coordinates with $\mu=0\dots d-1$.  
It will be assumed that $\phi,h_{\mu\nu},A_\mu$ do not depend on the
coordinate $u$. This 
admits the interpretation of $u$ as a coordinate on an $S^1$, that
is, subject to periodic identification. 

Let us consider zero-modes of strings in this background. The zero-mode
ansatz reads: 
\bel{zmans}
(X^A(\sigma,\tau))  = (t(\tau), x^1(\tau), \dots, 
x^{d-1}(\tau), w\sigma) \ . 
\ee
This class of configurations describes ground states of a string winding
$w$ times 
around the direction 
$u\equiv x^d$. The Nambu-Goto action 
for these modes reduces to
\bel{kkkpartact}
S_0 = -\int d\tau\ m \sqrt{- h_{\mu\nu} \dot{x}^\mu \dot{x}^\nu} \ ,
\ee
where now
\bel{kkmass}
m = \frac{|w|}{2\alp} e^{\phi} \ .
\ee
This has an obvious interpretation. From the point of view of winding
string zero-modes only the spacetime transverse to the cycle is visible. 
The Nambu-Goto action reduces to that of a
particle with mass given by 
\rf{kkmass} moving in the geometry described by the metric
$ds_\perp^2$. This mass is just the string tension multiplied by the proper
length of the winding string. In the spirit of \cite{Turok:2004gb} one can
make this argument more 
formally in 
the Hamiltonian formalism -- this is sketched in the appendix.

Suppose now that $\phi$ depends only on time and that $h_{\mu\nu}$ is
static. In that case, at least for  
slowly varying $\phi$, it is natural to regard $R(t)\equiv e^\phi$ as the
radius of the compact direction. If the original spacetime involves a cycle
which shrinks to zero and re-expands (for example, if
$\phi\sim\gamma\ln(|t|)$ for positive $\gamma$), then the situation is
similar to the Milne example.  

Equivalently, one can regard 
\rf{kkkpartact} as the action of a massive particle in a spacetime
described by the metric 
\be\label{kkredmetric}
ds^2 = e^{2\phi} h_{\mu\nu} dx^\mu dx^\nu \ .
\ee
These two possibilities clearly correspond to the freedom of choosing 
either the Einstein frame or the string frame in the field theory low
energy effective action. 
The metric \rf{kkredmetric}
has a real 
singularity under the above assumptions, but (as discussed in the previous
section) one expects that there
should be smooth 
geodesics 
passing through the singular locus. 
While it may
be hard to find geodesics for a specific metric of the form 
\rf{kkredmetric} explicitly, in 
many cases one may confirm this expectation by computing the
Hamiltonian and verifying that it is not singular.

\section{Examples}
\label{sect:examples}

This section is devoted to some simple examples. First consider 
spacetimes of the form 
\be
ds^2 = -dt^2 + \beta^2 t^{2n} d\theta^2 + d\vec{x}^2 \ , 
\ee
for integers $n>1$. In this case there is a curvature singularity
regardless of the periodicity of $\theta$. The discussion of winding 
strings can be carried out just as 
for the Milne case discussed in section \rf{sect:milne}, the only
modification  
being that the effective mass \rf{milmass} has to be replaced by
\bel{efmas}
m(t) = \frac{|w|\beta}{2\alp} |t|^n \ .
\ee
The equation of motion \rf{eom} can also be integrated analytically, giving
a smooth 
and unique answer which can be expressed in terms of elliptic functions. 
It is amusing to note that defining the ``cosmic time'' $\xi=t^n$ one has a
Friedman-Robertson-Walker spacetime with
$a(\xi)=((n+1)\xi)^{\frac{n}{n+1}}$ which corresponds to 
the equation of state $p=\kappa\rho$ with $\kappa=(2-n)/3n$. Thus the
case $n=2$ 
corresponds to pressureless dust. 

Another simple example is
the metric considered by a number of authors (see e.g. \cite{Craps:2002ii})
as an analytic
continuation of Witten's two-dimensional black hole \cite{Witten:1991yr}:
\be\label{aaa}
ds^2 = -dt^2 + \beta^2 \tanh^2 t \ d\theta^2 + d\vec{x}^2 \ .
\ee
The metric \rf{aaa} describes a curved spacetime: unlike the locally 
flat Milne
example, in this case the 
Riemann tensor is nontrivial. Similarly to the Milne case the locus $t=0$
is not singular unless $\theta$ is compactified: one finds, e.g. that 
\be
R_{\mu\nu\lambda\rho} R^{\mu\nu\lambda\rho} = 10\ {\mbox{\rm sech}}^4 t \ , 
\ee
which vanishes at $t=0$. However, when the coordinate $\theta$ is
periodically 
identified so as to describe a compact dimension, there
is a conical singularity similar to what occurs for the Milne
spacetime. 
The Nambu-Goto action for zero-modes of the form \rf{zmans} 
in synchronous gauge reduces to the same for as \rf{milneact} 
but with 
\be
m(t) = \frac{\beta}{2\alp} |w\tanh t| \ .
\ee
In this case it is also simple to integrate the equation of motion \rf{eom}  
analytically, and one finds a unique and nonsingular result. This is
perhaps not surprising, since 
the metric \rf{aaa} reduces to Milne at small $t$.

\section{Multiple cycles}
\label{sect:multiple}

It is natural to ask whether the arguments given in the previous sections
generalize to 
situations where there are multiple shrinking cycles. 
On physical grounds one 
expects that any non-vanishing momentum in those directions will lead to
divergences as the cycles shrink. Thus, even if an extended objects is
wrapping one 
shrinking cycle it will still have singular evolution if it can move in
another direction which is shrinking. At the very least, one can say that
the system will be unstable. 
It might seem at first glance that 
if all shrinking directions
were wrapped, then blue-shift singularities would not appear, 
by a directly generalizing the arguments of Turok et
al. \cite{Turok:2004gb}. Whether this can work or not depends on whether
there is enough gauge symmetry to render motion the wrapped dimensions
unphysical (``pure gauge''). In the case of wrapped strings it is easy to
see that already in the case of two cycles shrinking at the same time there
is a 
linear combination of them which is ``pure gauge'' in the above sense, and
an orthogonal direction which is not. Thus the system is unstable aganist
blue-shift divergences once there is any momentum in the latter
dimension. So for strings one can have zero-modes smoothly crossing the
singular locus only in the case of a single shrinking dimension. 

One can verify this explicitly in a simple class of spacetimes generalizing
\rf{kkansatz}, namely 
\bel{mkkansatz}
ds^2 = \sum_{i=1}^p e^{2\phi_i} (du_i + A^{(i)})^2 + ds_\perp^2 \ ,
\ee
where one has $p$ commuting isometries along the directions parameterized
by $u_i$ and 
\be
ds_\perp^2 = h_{\mu\nu} dx^\mu dx^\nu, \qquad A^{(i)} = A^{(i)}_\mu dx^\mu
\ . 
\ee
Here $\phi_i,h_{\mu\nu}, A^{(i)}_\mu$ depend only on the ``transverse''
coordinates   
$x_\perp\equiv (x^1,\dots,x^{d-p})$ and possibly on $t$. Thus 
one can interpret the $u^i$ as being subject to periodic
identification\footnote{All periodic coordinates are taken to have a fixed
periodicity $2\pi$.}.   
For simplicity, suppose also that the ``dilatons'' $\phi_i$ depend only on
time. A specific example would be a string winding a two-dimensional torus
with time-dependent radia: 
\be
ds^2 = -dt^2 + t^2 (\beta_1^2 (dx^d)^2 + \beta_2^2 (dx^{d-1})^2) +
ds_\perp^2 \ . 
\ee
This is a simple generalization of the Milne example where two torus cycles
shrink to zero radius and then expand at the same rate. 

Suppose $p$ cycles are wrapped by the string. In such a situation the
zero-mode ansatz generalizing \rf{zmans} takes the form 
\bel{multwind}
(X^A(\sigma,\tau))  = (t(\tau), x^1(\tau), \dots, x^{d-p}(\tau), 
w_1\sigma,\dots, w_p\sigma) \ ,
\ee
where $w_1\dots w_p$ are winding numbers. This class of configurations
describes ground states of a string winding 
around the isometry directions. 
The Nambu-Goto action \rf{nambugoto} for these modes reduces to
\be\label{mkkpartact}
S_0 = -\int d\tau\ m \sqrt{ - h_{\mu\nu} \dot{x}^\mu \dot{x}^\nu} \ ,
\ee
where now
\bel{mulmass}
m = \frac{1}{2\alp} \sqrt{\sum_i w_i^2 e^{2\phi_i}} \ .
\ee
The interpretation is that winding
strings move in the transverse space
and their mass is determined by their proper length. 
The Hamiltonian is given by \rf{milham} with the time-dependent mass given
by \rf{mulmass}. It would seem at this point that there is no blue-shift
divergence. 
The problem with this reasoning is 
that already the Ansatz \rf{multwind} does not allow for any motion in the
isometry 
directions. 
If one allows for any
nonvanishing momentum in these directions the blue-shift divergence
will appear. There is no argument generalizing that of
\cite{Turok:2004gb} 
that can prevent this from happening. 
The problem is not apparent, because 
in formula \rf{multwind} there 
are no  $\tau$ dependent terms in the isometry directions. For the case of
of a single shrinking dimension this was justified by 
the analysis performed in \cite{Turok:2004gb}, which pointed out that for
strings uniformly winding this dimension motion in this direction is not
physical. This no longer suffices if 
there is more than one shrinking cycle. This can be verified by analysing the
equations of motion which follow from the Nambu-Goto action. Equivalently,
one can analyse them starting with the Polyakov action \rf{polyakov} and 
the Virasoro constraints.
%For the familiar 
%case of a string (as opposed to higher $p$-branes) 
It is easy to see that
there is always only one linear combination of momenta which is set to zero
without instability. In the Polyakov approach this direction is given by 
the linear combination of cycles defined by zero-modes of the Virasoro
constraints\footnote{Specifically, this linear combination can be read off
  from the level-matching constraint
$\bar{L}_0=L_0$ (using standard notation).}. The directions orthogonal to
this dimension do however suffer from a 
blue-shift instability. A systematic analysis could also be carried out in
the Hamiltonian formalism, as done by Turok et
al. \cite{Turok:2004gb} for the case of Milne space. Clearly, for the case
of higher $p$-branes there 
will also be limits to how many cycles can shrink without necessarily
leading to instability.

\section{Conclusions}

It was noted in \cite{Turok:2004gb} that at the classical level extended
objects winding the 
shrinking circle in Milne spacetime evolve smoothly through the singularity. 
This note discussed the extension of the arguments given there to some more
general examples.  

The main focus of \cite{Turok:2004gb} was the physics of winding membrane
zero-modes. The discussion of M-theory membranes in 
Milne spacetime presented there generalizes to some of the more general
spacetimes 
discussed above (i.e. \rf{kkansatz} and \rf{mkkansatz}) in 
the case of winding strings. However when more than one cycle shrinks there
are limits on the dimensionality of extended objects which evolve smoothly
and are stable. Specifically, in case of winding strings such stable and
smooth evolution  (along the lines of \cite{Turok:2004gb} is possible only
if no more than one cycle is shrinking.  

The winding string examples discussed earlier in this paper can be viewed
as membranes multiply wrapped on tori with some of the torus cycles
degenerating in time. Instead of the Nambu-Goto action for the string one
begins with the membrane action and an appropriate zero-mode ansatz. This
leads to a nonsingular Hamiltonian if all shrinking dimensions are
wrapped (up to the limitations discussed in section
\ref{sect:multiple}). The effective description thus obtained pertains to a
string if only one membrane dimension is wrapped, or a particle if two
membrane dimensions are wrapped. 

The arguments presented here were based on two approximations: they
were limited to the classical approximation, and furthermore to a
minisuperspace 
approximation which ignored all non-zero string modes. To draw firm
conclusions 
about the physics of this problem it would be crucial to understand 
the validity of this procedure. 
This would require 
a proper quantum treatment, taking into account all worldsheet degrees of 
freedom, not just the embedding coordinates $X$. 
It is clearly
fascinating to pursue these questions and understand what message can be
inferred from the apparently smooth passage enjoyed by winding extended
objects in singular spacetimes.

\medskip
\centerline{{\bf\large Acknowledgements}}

The authors would like to thank Robert Budzy\'nski, Andrzej
Krasi\'nski and Jerzy Lewandowski for helpful discussions.

\newpage

\appendix

\section{Appendix: Hamiltonian reduction of winding string dynamics}
\label{Appendix}
\renewcommand{\theequation}{A.\arabic{equation}}
\setcounter{equation}{0}

This section presents a more formal argument showing that 
the metrics \rf{kkansatz} lead to nonsingular evolution. This is a simple
generalization of the analysis carried out by Turok et
al. \cite{Turok:2004gb}.  
The starting point is the Polyakov action which 
has the form
\begin{equation}\label{polyaction}
S=\int d\tau L(\tau),~~~~L(\tau):=-\frac{\mu_1}{2}\int d\sigma
\sqrt{-\gamma}\:\gamma^{ab}\partial_a X^A \partial_b X^B g_{AB},
\end{equation}
where $(X^A)\equiv (X^0,X^1,\ldots,X^{d-1}, \Theta)\equiv
(X^0,X^k,\Theta)\equiv (X^\mu, \Theta) $ are embedding functions of a
string in spacetime with metric $g_{AB}$, $~\gamma_{ab}$ is the
intrinsic metric of the string worldsheet ($\gamma$ is its
determinant), and $(a)\equiv (\tau,\sigma)$.

Owing to reparameterization invariance of the action \rf{polyaction} 
the system  has constraints. Detailed analysis of the corresponding
constrained dynamics has been carried out by Turok et al. The final
form of the corresponding Hamiltonian reads
\begin{equation}\label{ham1}
    H= \int d\sigma (\frac{1}{2}\tilde{A}\tilde{C} + \tilde{A^1} \tilde{C_1}) ,
\end{equation}
where $\tilde{A}$ and $\tilde{A^1}$ are arbitrary functions of
$\tau$ and $\sigma$, and where
\begin{equation}\label{con1}
    \tilde{C}:=\Pi_A \Pi_B g^{AB}+ \mu_1^2\acute{X}^A \acute{X}^B
    g_{AB} =0, ~~~~~~\tilde{C_1}:= \Pi_A \acute{X}^A =0 ,
\end{equation}
are first-class constraints ($\Pi_A:= \partial L
/\partial\dot{X}^A,~\dot{X}^A :=\partial X^A/\partial \tau,~
\acute{X}^A:=\partial X^A/\partial \sigma$).

In case of a string twisted around $u$-direction defined by (3.3) and
in the gauge $\tilde{A^1}=0$, Hamilton's equations lead to the new
constraint $\Pi_U =0$. The set of constraints for $\tilde{C}$,
$U-w\sigma$ and $\Pi_U $ is not first-class because the Poisson
bracket between $\tilde{C}$ and each of the other two constraints
does not vanish. The solution is treating  $U=w\sigma$ and $\Pi_U =0$
as the second-class constraints. In that scheme the Dirac bracket
amounts to canceling the $U$ and $\Pi_U$ derivatives from the Poison
bracket. This way one can eliminate the variables $U$ and $\Pi_U$
from the dynamics. As the result the dynamics of a string in $d+1$
dimensional spacetime reduces to the dynamics of a `particle' in
spacetime with the dimension $d$. The corresponding Hamiltonian can
be obtained by substituting $U=w\sigma$ and $\Pi_U =0$ into
\rf{con1}. Finally, one gets
\begin{equation}\label{ham2}
H=\tilde{A}(\tau)\tilde{C}=
    \tilde{A}(\tau)\:(\Pi_\mu\Pi_\nu\tilde{h}^{\mu\nu}+\mu_1^2 
    w^2 e^{2\phi}) \ ,
\end{equation}
where $\tilde{h}^{\mu\nu}$ is the inverse of the metric
$\tilde{h}_{\mu\nu}= h_{\mu\nu}+A_\mu A_\nu e^{2\phi}$ defined by
(3.1).

It results from \rf{ham2} that a string twisted around the
direction $u$, behaves like a particle with an effective mass $~\mu_1
w e^\phi~$ in spacetime with metric $\tilde{h}_{\mu\nu}$. The
character of dynamics defined by the Hamiltonian \rf{ham2} depends
on regular/singular properties of $\tilde{h}_{\mu\nu}$.


\newpage

\begin{thebibliography}{00}



%\cite{Dixon:1985jw}
\bibitem{Dixon:1985jw}
  L.~J.~Dixon, J.~A.~Harvey, C.~Vafa and E.~Witten,
  ``Strings On Orbifolds,''
  Nucl.\ Phys.\ B {\bf 261}, 678 (1985).
  %%CITATION = NUPHA,B261,678;%%

%\cite{Gasperini:2002bn}
\bibitem{Gasperini:2002bn}
  M.~Gasperini and G.~Veneziano,
  ``The pre-big bang scenario in string cosmology,''
  Phys.\ Rept.\  {\bf 373}, 1 (2003)
  [arXiv:hep-th/0207130].
  %%CITATION = HEP-TH 0207130;%%


%\cite{Horowitz:ap}
\bibitem{Horowitz:ap}
  G.~T.~Horowitz and A.~R.~Steif, 
  ``Singular String Solutions With Nonsingular Initial Data,'' 
  Phys.\ Lett.\ B {\bf 258}, 91 (1991).
%%CITATION = PHLTA,B258,91;%%

\bibitem{Khoury:2001bz}
  J.~Khoury, B.~A.~Ovrut, N.~Seiberg, P.~J.~Steinhardt and N.~Turok,
  ``From big crunch to big bang,''
  Phys.\ Rev.\ D {\bf 65}, 086007 (2002)
  [arXiv:hep-th/0108187].
  %%CITATION = HEP-TH 0108187;%%

%\cite{Seiberg:2002hr}
\bibitem{Seiberg:2002hr}
  N.~Seiberg, 
  ``From big crunch to big bang - is it possible?,'' 
  arXiv:hep-th/0201039.
  %%CITATION = HEP-TH 0201039;%%


%\cite{Cornalba:2002fi}
\bibitem{Cornalba:2002fi}
  L.~Cornalba and M.~S.~Costa,
  ``A new cosmological scenario in string theory,''
  Phys.\ Rev.\ D {\bf 66}, 066001 (2002)
  [arXiv:hep-th/0203031].
  %%CITATION = HEP-TH 0203031;%%


%\cite{Nekrasov:2002kf}
\bibitem{Nekrasov:2002kf}
  N.~A.~Nekrasov, 
  ``Milne universe, tachyons, and quantum group,'' 
  arXiv:hep-th/0203112.
  %%CITATION = HEP-TH 0203112;%%



%\cite{Liu:2002yd}
\bibitem{Liu:2002yd}
  H.~Liu, G.~W.~Moore and N.~Seiberg,
  ``The challenging cosmic singularity,''
  arXiv:gr-qc/0301001.
  %%CITATION = GR-QC 0301001;%%


%\cite{Pioline:2003bs}
\bibitem{Pioline:2003bs}
  M.~Berkooz, and B.~Pioline,
  ``Strings in an electric field, and the Milne universe,''
  JCAP {\bf 0311} (2003) 007
  [arXiv:hep-th/0307280].
  %%CITATION = HEP-TH 0307280;%%


\bibitem{Berkooz:2004re}
  M.~Berkooz, B.~Pioline and M.~Rozali,
  ``Closed strings in Misner space: Cosmological production of winding
  strings,'' 
  JCAP {\bf 07} (2004) 003 [arXiv:hep-th/0405126].
  %%CITATION = HEP-TH 0405126;%%

%\cite{Berkooz:2005ym}
\bibitem{Berkooz:2005ym}
  M.~Berkooz, Z.~Komargodski, D.~Reichmann and V.~Shpitalnik,
  ``Flow of geometries and instantons on the null orbifold,''
  arXiv:hep-th/0507067.
  %%CITATION = HEP-TH 0507067;%%


%\cite{Cornalba:2003kd}
\bibitem{Cornalba:2003kd}
  L.~Cornalba and M.~S.~Costa,
  ``Time-dependent orbifolds and string cosmology,''
  Fortsch.\ Phys.\  {\bf 52}, 145 (2004)
  [arXiv:hep-th/0310099].
  %%CITATION = HEP-TH 0310099;%%

%\cite{Durin:2005ix}
\bibitem{Durin:2005ix}
  B.~Durin and B.~Pioline,
  ``Closed strings in Misner space: A toy model for a big bounce?,''
  arXiv:hep-th/0501145.
  %%CITATION = HEP-TH 0501145;%%

%\cite{Gubser:2003vk}
\bibitem{Gubser:2003vk}
  S.~S.~Gubser,
  ``String production at the level of effective field theory,''
  Phys.\ Rev.\ D {\bf 69}, 123507 (2004)
  [arXiv:hep-th/0305099].
  %%CITATION = HEP-TH 0305099;%%


%\cite{Turok:2004yx}
\bibitem{Turok:2004yx}
  N.~Turok and P.~J.~Seinhardt,
  ``Beyond inflation: A cyclic universe scenario,''
  Phys.\ Scripta {\bf T117}, 76 (2005)
  [arXiv:hep-th/0403020].
  %%CITATION = HEP-TH 0403020;%%

%\cite{Steinhardt:2004gk}
\bibitem{Steinhardt:2004gk}
  P.~J.~Steinhardt and N.~Turok,
  ``The cyclic model simplified,''
  New Astron.\ Rev.\  {\bf 49}, 43 (2005)
  [arXiv:astro-ph/0404480].
  %%CITATION = ASTRO-PH 0404480;%%


%\cite{Horava:1996ma}
\bibitem{Horava:1996ma}
  P.~Horava and E.~Witten,
  ``Eleven-Dimensional Supergravity on a Manifold with Boundary,''
  Nucl.\ Phys.\ B {\bf 475}, 94 (1996)
  [arXiv:hep-th/9603142].
  %%CITATION = HEP-TH 9603142;%%

%\cite{Horava:1995qa}
\bibitem{Horava:1995qa}
  P.~Horava and E.~Witten,
  ``Heterotic and type I string dynamics from eleven dimensions,''
  Nucl.\ Phys.\ B {\bf 460}, 506 (1996)
  [arXiv:hep-th/9510209].
  %%CITATION = HEP-TH 9510209;%%



%\cite{Turok:2004gb}
\bibitem{Turok:2004gb}
  N.~Turok, M.~Perry and P.~J.~Steinhardt,
  ``M theory model of a big crunch / big bang transition,''
  Phys.\ Rev.\ D {\bf 70}, 106004 (2004)
  [Erratum-ibid.\ D {\bf 71}, 029901 (2005)]
  [arXiv:hep-th/0408083].
  %%CITATION = HEP-TH 0408083;%%
 
%\cite{Townsend:2003fx}
\bibitem{Townsend:2003fx}
  P.~K.~Townsend and M.~N.~R.~Wohlfarth,
  ``Accelerating cosmologies from compactification,''
  Phys.\ Rev.\ Lett.\  {\bf 91}, 061302 (2003)
  [arXiv:hep-th/0303097].
  %%CITATION = HEP-TH 0303097;%%


%\cite{Ohta:2003pu}
\bibitem{Ohta:2003pu}
  N.~Ohta,
  ``Accelerating cosmologies from S-branes,''
  Phys.\ Rev.\ Lett.\  {\bf 91}, 061303 (2003)
  [arXiv:hep-th/0303238].
  %%CITATION = HEP-TH 0303238;%%

%\cite{Ohta:2003ie}
\bibitem{Ohta:2003ie}
  N.~Ohta,
  ``A study of accelerating cosmologies from superstring / M theories,''
  Prog.\ Theor.\ Phys.\  {\bf 110}, 269 (2003)
  [arXiv:hep-th/0304172].
  %%CITATION = HEP-TH 0304172;%%


%\cite{Chen:2003dc}
\bibitem{Chen:2003dc}
  C.~M.~Chen, P.~M.~Ho, I.~P.~Neupane, N.~Ohta and J.~E.~Wang,
  ``Hyperbolic space cosmologies,''
  JHEP {\bf 0310}, 058 (2003)
  [arXiv:hep-th/0306291].
  %%CITATION = HEP-TH 0306291;%%


%\cite{Ohta:2004wk}
\bibitem{Ohta:2004wk}
  N.~Ohta,
  ``Accelerating cosmologies and inflation from M / superstring theories,''
  Int.\ J.\ Mod.\ Phys.\ A {\bf 20}, 1 (2005)
  [arXiv:hep-th/0411230].
  %%CITATION = HEP-TH 0411230;%%


%\cite{Brandenberger:1988aj}
\bibitem{Brandenberger:1988aj}
  R.~H.~Brandenberger and C.~Vafa,
  ``Superstrings In The Early Universe,''
  Nucl.\ Phys.\ B {\bf 316}, 391 (1989).
  %%CITATION = NUPHA,B316,391;%%

%\cite{Costa:2005ej}
\bibitem{Costa:2005ej}
  M.~S.~Costa, C.~A.~R.~Herdeiro, J.~Penedones and N.~Sousa,
  ``Hagedorn transition and chronology protection in string theory,''
  arXiv:hep-th/0504102.
  %%CITATION = HEP-TH 0504102;%%

%\cite{Tolley:2005us}
\bibitem{Tolley:2005us}
  A.~J.~Tolley,
  ``String propagation through a big crunch / big bang transition,''
  arXiv:hep-th/0505158.
  %%CITATION = HEP-TH 0505158;%%

%\cite{McGreevy:2005ci}
\bibitem{McGreevy:2005ci}
  J.~McGreevy and E.~Silverstein,
  ``The tachyon at the end of the universe,''
  arXiv:hep-th/0506130.
  %%CITATION = HEP-TH 0506130;%%


%\cite{Craps:2002ii}
\bibitem{Craps:2002ii}
  B.~Craps, D.~Kutasov and G.~Rajesh,
  ``String propagation in the presence of cosmological singularities,''
  JHEP {\bf 0206}, 053 (2002)
  [arXiv:hep-th/0205101].
  %%CITATION = HEP-TH 0205101;%%

%\cite{Witten:1991yr}
\bibitem{Witten:1991yr}
  E.~Witten,
  ``On string theory and black holes,''
  Phys.\ Rev.\ D {\bf 44}, 314 (1991).
  %%CITATION = PHRVA,D44,314;%%

\bibitem{luth}
  D. Lust and S. Theisen, 
  ``Lectures in String Theory'',
  Lecture Notes Phys. 346 (Springer-Verlag, Berlin, 1989).

\bibitem{jpol} J.Polchinski, 
  ``String Theory'' (Cambridge Univer. Press, 1998), Vols. 1 and 2. 



%%%%%%%%%%%%%%%%%%%%%%%%% ????


\end{thebibliography}
\end{document}